\documentclass[11pt]{article} 

\begin{document}

\begin{flushright}
                                                   IITK-PHY-99-51 \\
                                                   hep-ph/9905576 \\
\end{flushright} 
\begin{center}
{\large\bf  TRILINEAR R-PARITY VIOLATION: THEORY TO EXPERIMENT }
\vskip 5pt
{\large\sl               Sreerup Raychaudhuri}
\vskip 5pt
{\rm
Department of Physics, 
Indian Institute of Technology, 
Kanpur 208 016, India.} \\
{\rm E-mail: sreerup@iitk.ac.in }
\end{center} 
\begin{center}
{\sc Abstract} 
\end{center}
{\footnotesize\rm
Supersymmetric models without conservation of $R$-parity are reviewed and
low-energy constraints on the extra trilinear couplings listed.  Current 
searches at the LEP, Tevatron and HERA colliders are then summed up. 
Prospects for further study, especially at future colliders, are briefly 
touched upon. 
}
\vskip 5pt
\noindent
{\footnotesize\it  Expanded Version of an Invited Talk delivered at the 
13$^{th}$ Topical Conference on Hadron Collider Physics, Mumbai, 
India (January 1999). }

\bigskip\bigskip

\centerline{\large\sc 1. 
Introduction to R-Parity and R-Parity Violation} 
\bigskip

As we enter a new millenium, supersymmetry (SUSY) appears to be one of the 
best options for physics beyond the Standard Model (SM). One of the 
cornerstones of 
conventional searches for SUSY is the idea of a conserved quantum number 
called $R$-parity, which leads to missing energy-momentum triggers for SUSY 
signals. We first examine the rationale for conserving $R$-parity. 

In the minimal supersymmetric extension of the SM, or MSSM, as it is 
commonly dubbed, to each particle $P$ in the SM, there corresponds 
a {\it super}field 
$\hat P$ which contains the field $P(x)$, its SUSY partner (sparticle) field
${\widetilde P}(x)$ and an auxiliary field $F^P(x)$. The last of these
has no dynamics and can be easily integrated out of the action.
The simplest prescription for obtaining (most of) the interactions 
of the matter sector of the MSSM is to take the terms in the SM Lagrangian
and replace each field $P(x)$ by the corresponding superfield $\hat P$. 
This leads to the generation of a superpotential, from which the interaction
vertices can be obtained by standard rules of composition.
This results in the well-known prescription for generating the vertices
of the MSSM, namely, to take each SM vertex 
and replace a {\it pair} of SM particles by their SUSY partners (sparticles). 
This prescription is equivalent to postulating\cite{Farrar-Fayet} 
conservation of a multiplicative quantum number, namely $R$-parity, which 
is defined by 
$$ R_p = (-1)^{3B+L+2S} \ ,$$ 
where $B,L,S$ denote, respectively, the baryon number, lepton number and 
intrinsic spin of the particle. Each SM particle has $R_p~=~+1$, while
sparticles generically have $R_p~=~-1$. Conservation of $R_p$ is not really 
a surprising feature, since the MSSM interactions constructed by 
the method described above may be expected to retain the symmetries of 
the SM, where, 
as is well-known, the quantum numbers $B$ and $L$ are conserved separately. 
It is natural, therefore, to expect the same to hold for the MSSM. This
immediately leads us to conservation of $R$-parity.

$R$-parity conservation, in fact, turns out to be a very convenient 
feature of SUSY models. For one thing, it means that sparticles are always 
produced in pairs and hence the {\it lightest} sparticle (LSP) must be stable.
The LSP will actually behave like a heavy neutrino, since it can only interact 
with ordinary matter by exchanging other (heavy) sparticles. Once a sparticle
has been produced in some high energy process, all cascades 
resulting from its decay must end in production of an LSP, 
which (like a neutrino) escapes the detectors, generating characteristic 
missing energy signals. This facilitates experimental detection, especially
at hadron colliders, by providing a simple trigger to base the searches on. 
For this reason, over the years, a considerable literature has grown
up around these missing energy signals, and
the same ideas have been extensively used in experimental searches for 
SUSY.

Another good feature of conserved $R$-parity is that 
the LSP turns out to be an excellent candidate for the cold dark matter 
component of the Universe. This is demanded by theoretical models which 
close the Universe and especially by theories which postulate an inflationary
epoch in the early Universe. 
Observational evidence, such as the rotation curves of spiral galaxies, 
does seem to require some dark matter, and the observed level of fluctuations
in the microwave background also requires a cold dark matter component in
order to explain galaxy formation. However, it is only fair to point out
that there could be other candidates for cold dark matter, such as invisible
axions, other wimps or machos.  \\ 

{\sl 1.1 ~~~\underline{Violation of R-Parity}}  \\

The story of $R$-parity conservation does not end here, however. It is
surely not reasonable to impose a dynamical symmetry merely because it has
agreeable consequences. We must therefore, ask ourselves if the arguments
leading to the postulate of $R$-parity conservation are really compelling. 
It turns out that 
the weak point of the above scenario is that $B$ and $L$ are, in a sense, 
accidental symmetries of the SM and are not built into the gauge symmetry. 
Hence, the moment one seeks to extend the SM ({\it e.g.}, to get a GUT), 
the possibility of $B$ and/or $L$-violation must be encountered. The MSSM 
is just one case of this\cite{Weinberg}. Here the 
source of $L$-violation lies in 
the well-known fact that there are {\it two} Higgs doublets instead of
one, as in the SM. This is forced upon us by the requirement of 
holomorphicity of the superpotential as well as arguments from 
anomaly cancellation: these arguments also tell us that the two doublets
must have opposite hypercharges. Thus, there must be a scalar 
doublet superfield $\hat{H}_2$ with hypercharge $Y = -1$ in addition to 
the SM-like $\hat{H}_1$, which has $Y = +1$. It is obvious that the 
$\hat{H}_2$ has gauge quantum numbers identical to those of the left-handed 
lepton superfields $\hat{L}_i$ and hence can {\it mix} with all of them. 
Accordingly, to the $R$-parity conserving superpotential of the MSSM, we 
can always add terms with the $\hat{H}_2$ replaced by a generic lepton
superfield $\hat{L}_i$ and some unknown coupling(s), 
\begin{eqnarray}
{\cal W} & = &  + \mu \hat H_1 \hat H_2
                + h^L_{ik} \hat L_i \hat H_2 \hat E^c_k
                + h^D_{ik} \hat Q_i \hat H_2 \hat D^c_k
                + \dots   \nonumber \\
         &   &  + \kappa_i \hat H_1 \hat L_i
                + \lambda_{ijk} \hat L_i \hat L_j \hat E^c_k
                + \lambda'_{ijk} \hat Q_i \hat L_j \hat D^c_k \nonumber \\
         &   &  + \lambda''_{ijk} \hat U^c_i \hat D^c_j \hat D^c_k \ . 
\label{superpot}
\end{eqnarray}

It is also possible to violate baryon number by multiplying three right-handed
$SU(2)_L$-singlet quark superfields together and this gives the last term in 
Eq.~(\ref{superpot}). In the above equation, the indices $i,j,k$ represent 
generations of SM fermions. 
The condition that the superpotential be holomorphic in 
the fields requires that the $SU(2)_L$ doublets be multiplied using the 
$SU(2)$ product $\epsilon_{ab} \Phi_1^a \Phi_2^b$ rather than the simpler 
$\Phi_1^\dagger \Phi_2$. This ensures that the couplings $\lambda_{ijk}$ are 
antisymmetric in $i,j$. Similarly since the quark superfields belong 
to {\bf 3} of $SU(3)_c$, the term in the Lagrangian must be the singlet in 
the decomposition of {\bf 3}$\times${\bf 3}$\times${\bf 3}, {\it i.e.} 
totally antisymmetric in colour indices. This ensures that the 
$\lambda''_{ijk}$ are also antisymmetric in the flavour indices $j,k$. 

At first sight it seems as if the bilinear terms 
$\kappa_i \hat H_1 \hat L_i$
can be rotated away leaving only the trilinear ones. 
This is, in fact, stated in several (early) works on the subject.  
However, this rotation would mean that the Higgs superfield $\hat{H}_2$ 
acquires
an admixture of leptonic superfields and this will show up in the 
scalar potential of the theory. It is then possible for the 
sneutrino\footnote{like the neutral Higgs boson, with which, indeed, it 
mixes} to acquire a vacuum expectation value (VEV), leading to spontaneous 
violation of $R$-parity. In such scenarios, the charginos mix with the
charged leptons while the neutralinos mix with the neutrinos. Similar
mixings occur among the scalars of the theory. Even if the sneutrino 
does not develop a VEV, it is possible to have bilinear $R$-parity
violation on the same footing as trilinear $R$-parity violation. 
Moreover, bilinear terms induce the trilinear terms (with a
definite structure) through
the Higgs Yukawa couplings and hence, in a sense, such models are more
predictive than the ones discussed in this article.
There exists a rich
literature\cite{Hall-Suzuki} on the subject of bilinear and spontaneous
$R$-parity violation which is well worth perusal. Phenomenological 
aspects of these scenarios require a detailed consideration, but we 
shall not attempt to discuss them in this article. 

It is also worth mentioning that the above interactions tell us that in the
presence of $\lambda$ couplings, the sleptons behave as {\it dileptons};
in the presence of $\lambda'$ couplings, the squarks behave as
{\it leptoquarks}; in the presence of $\lambda''$ couplings, the squarks
behave as {\it diquarks}. Such objects at the electroweak scale have been 
predicted in composite models, but here it is elementary particles which
exhibit similar behaviour. Thus any search for scalar 
dileptons, diquarks or leptoquarks becomes automatically
applicable to $R$-parity-violating effects. Vector
excitations of this nature cannot be mapped to any SUSY effects and,
therefore, are irrelevant for our purposes.  \\

{\sl 1.2 ~~~\underline{Proton Decay}} \\ 

One of the immediate consequences of simultaneous
violation of $B$ and $L$ is known to 
be fast proton decay. Most Grand Unified Theories (GUT's) which 
try to unify the strong and electromagnetic interactions within a single 
(broken) gauge symmetry do predict proton decay as a consequence. In the MSSM
too, if $R$-parity is violated in a maximal sense, {\it i.e.}, 
all the extra 
couplings in Eq.(\ref{superpot}) are present, one can have proton decay 
through diagrams such as the one in Figure 1. 

\vskip 5pt
\thicklines
\begin{picture}(250,50)(-50,10)
\setlength{\unitlength}{0.6pt}

\put(100,80){\line(1,0){300}}
\put( 50,10){\line(1,0){350}}
\put( 50, 0){\line(1,0){350}}
\put(100,80){\line(-1,-2){30}} 
\put( 50,20){\line(1,0){21}}

\multiput(250,75)(0,-10){7}{\line(0,1){5}}

\put(250,80){\circle*{6}}
\put(250,10){\circle*{6}}

\put(  0,0){\Large $p$}
\put(100,85){\large $u$}
\put(100,15){\large $u$}
\put(100,-20){\large $d$}
\put(400,-20){\large $d$}
\put(400,15){\large $\bar d$}
\put(400,85){\large $e^+$}
\put(265,45){\large $\widetilde d_R$}
\put(450,0){\Large $\pi^0$}
\put(425,0){\Huge \} } 
\put( 20,0){\Huge \{ }

\put(250,90){$\lambda'$}
\put(220,15){$\lambda''$}

\end{picture}

\vskip 25pt
\noindent {\bf Figure 1.} {\footnotesize\it 
Typical diagram for proton
decay $p \rightarrow e^+ \pi^0$ through $\lambda'$ and $\lambda''$
couplings.  
} 
\vskip 10pt

The amplitude for this process can immediately be estimated as 
\begin{equation}
{\cal A}~(p \rightarrow \ell^+ \pi^0) \sim 
\frac {\lambda' \lambda''}{M^2_{\widetilde d_R}} \ .
\end{equation}
Of course, this is not the only possible diagram, but all others 
have similar amplitudes. Unless one makes very special choices of
phase (which would then have to be explained), it is unlikely that 
there will be large cancellations between different (coherent)
diagrams and hence, one can make reasonable estimates of bounds
using the above result. 

Till date all experimental searches for proton decay have yielded
negative results, leading to a lower bound on the proton lifetime of $\sim
10^{32}$ years: this immediately constrains\cite{Goity-Sher} 
the product $\lambda' \lambda''$ in the above equation to be $\sim 10^{25}$
or smaller, for $M^2_{\widetilde d_R} < $ a few TeV. 
Rather than take on the burden of explaining such an unnaturally small
number, it is more reasonable to assume that the product 
$\lambda' \lambda''$ vanishes {\em identically}. 
Assuming this to be the case, if we wish this result to hold at 
all scales, it is clear that we must postulate some symmetry to protect it. 
$R$-parity is just such a symmetry, since it forbids both the
$\lambda'$ and the $\lambda''$ terms, but it is something of an 
overkill, since it is enough to have either of the factors vanish
{\it i.e.} we can either have $L$ conserved, with
vanishing of $\lambda'$, or have $B$ conserved, with vanishing $\lambda''$. 
In fact, since the early work of Ref.~\cite{Goity-Sher}, several
papers\cite{Smirnov-Vissani} 
have attempted to put bounds on all possible products of 
$R$-parity-violating couplngs which appear in various proton-decay modes. \\

{\sl 1.3 ~~~\underline{Discrete Symmetries}} \\

In the MSSM, $R$-parity is a relic of a global $U(1)$ symmetry called 
$R$-symmetry, which is broken to $Z_2$ when SUSY itself is broken. 
We can ensure conservation of $L$ by postulating {\it lepton parity}, 
another $Z_2$ symmetry under which all lepton superfields flip sign, 
while other superfields remain unchanged. Similarly, one can ensure 
conservation of $B$ by postulating {\it baryon parity}, under which all 
quark superfields change sign, while other superfields remain unchanged.
These parities must be imposed by hand in the 
MSSM\cite{Smirnov-Vissani} --- just as $R$-parity itself is --- and 
no compelling dynamical motivation has yet been discovered for either. 

The imposition of lepton or baryon-parity, however,
immediately leads one to ask  
the question: if the MSSM is the low-energy
effective theory of a SUSY GUT, then, at some high scale, 
the quarks and leptons must belong to the same gauge multiplet.
It is difficult, then, to visualise part of the multiplet being
even and part being odd under a discrete transformation such as lepton
or baryon parity. 

Two approaches to this paradox are possible. The (conceptually) 
simpler one is to assume that grand unification takes place within the 
framework of a string theory\cite{Hall-Ross} without any unified gauge 
group. In such models, one does not need to put quarks and leptons in 
the same multiplet, and hence there is no objection to having them
transform differently under a $Z_2$ symmetry . 
Even in the context of ordinary grand unification, 
however, it has been shown\cite{R-breaking} that it is possible to add 
$R$-parity conserving (nonrenormalisable)
operators to the Lagrangian of the theory; these, after the GUT
symmetry has been broken and quarks and leptons acquire separate 
identities, metamorphose into $R$-parity breaking (renormalisable)
operators. Thus, in these models, 
$R$-parity violation is possible without running foul
of proton decay constraints. 

It has also been argued that baryon parity is better motivated theoretically
than lepton parity. The reason is that it appears\cite{Krauss-Wilczek}
that quantum gravity
effects could maximally violate any discrete symmetry of a theory which
is not a relic of a gauged symmetry. At a high scale, then, we would 
require all the discrete symmetries of the Lagrangian to be such relics. An 
analysis\cite{Ibanez-Ross} 
of the possible candidate models shows that the only ones which are compatible
with an anomaly-free gauge theory, in the first place, are $R$-parity and
baryon parity. Accordingly, if we wish to break $R$-parity, it seems
that $L$ violation is favoured. However, one should note
that none of these arguments is really watertight and depend on various
approximations and {\em ans\"atze} which could be 
called into question\cite{Banks-Dine} if necessary.  

Finally, one is driven to the obvious question, as to whether 
$R$-parity-violating models are consistent with supersymmetric
coupling constant unification 
unification. The answer seems to be rather parameter-space 
dependent\cite{Allanach}, but it appears that low energy solutions of the 
renormalisation group in which $R$-parity violating couplings
partially drive the evolution are possible. In fact, these have been
used to put (somewhat loose) bounds on some of the $\lambda''$ 
couplings\cite{Brahmachari-Roy}. \\

\centerline{\large\sc 2. Low-Energy Phenomenology} 
\bigskip

Using the symmetry properties of the $R$-parity-violating couplings in 
Eq.(\ref{superpot}) it is possible to count the number of extra parameters 
introduced into the MSSM. There are 9 $\lambda_{ijk}$'s, 
27 $\lambda'_{ijk}$'s and 9 $\lambda''_{ijk}$'s, making a total of 45 
new parameters in all.  Some of these vanish because of the proton decay 
constraint --- for example, one can have 36 $\lambda_{ijk}$'s and
$\lambda'_{ijk}$'s, while the $\lambda''_{ijk}$'s vanish; or one can have 
9 $\lambda''_{ijk}$'s, while the rest vanish. Even then, 
enough free parameters remain to make  
a phenomenological analysis valueless if {\it all} of them are taken to be 
free parameters\footnote{This scenario has, in fact, been 
described\cite{Gunion} as a `nightmare'.}. 

In order to study the phenomenological behaviour of SUSY without 
$R$-parity conservation, then, the usual hypothesis one makes is that just one
of the $R$-parity-violating couplings is dominant and the others are, for 
all practical purposes,
zero. The common approach is to assume that this is true in the physical basis
for fermions. One can justify this by two plausibility arguments:
first, this is indeed the pattern of SM Yukawa couplings ($y_f$), where $y_t$ 
is much larger than the others; secondly, there exist various phenomenological
bounds on {\it products} of pairs of $R$-parity-violating
couplings\cite{Choudhury-Roy} which leads us to supect that these products 
might actually vanish. An alternative approach\cite{Agashe-Graesser} is
to assume that one coupling is dominant in the gauge basis (before the 
electroweak symmetry is broken) and others are generated because of the 
quark (lepton?) mixing 
in the physical basis through the Cabibbo-Kobayashi-Maskawa matrix in the 
quark (lepton?) sector. In the latter case, bounds on 
products of couplings translate into bounds on the single dominant coupling
in the gauge basis. This
latter procedure may seem a more rational one, since the former requires a
conspiracy among the couplings to have various cancellations in the
physical basis, leaving only one of them dominant. However, as some such 
mechanism is probably 
at work in the SM making $y_t$ large in the physical basis, the question is 
basically one of philosophy. We adopt the attutude that one of the couplings
in the physical basis is dominant, leaving the explanation to a deeper theory
when that should become available. \\

{\sl 2.1 ~~~\underline{Four-fermion operators}} \\

Assuming that one (at a time) out of the 45 possible couplings is dominant,
the Lagrangian arising out of Eq.(\ref{superpot})
leads\cite{Barger-Giudice-Han} to low-energy
effective four-fermion operators contributing to the physics of various 
processes at low and intermediate energy scales. These can be as
varied as \\
\hspace*{0.3in} $\bullet$ violation of charged current universality, \\
\hspace*{0.3in} $\bullet$ tau decays to final states with electrons/muons,\\
\hspace*{0.3in} $\bullet$ pion decays to electrons/muons, \\
\hspace*{0.3in} $\bullet$ generation of $\nu_e$ masses,\\
\hspace*{0.3in} $\bullet$ $\nu_\mu$ deep inelastic scattering,\\ 
\hspace*{0.3in} $\bullet$ neutrinoless double beta decay, \\
\hspace*{0.3in} $\bullet$ atomic parity violation, \\
\hspace*{0.3in} $\bullet$ $D^0-\bar D^0$ mixing,\\
\hspace*{0.3in} $\bullet$ $D^+$ decays, \\
\hspace*{0.3in} $\bullet$ leptonic decays of the $Z$, quantified by 
$R_\ell = \Gamma (Z \rightarrow hadrons) /
\Gamma (Z \rightarrow \ell^+\ell^-)$,\\
\hspace*{0.3in} $\bullet$ heavy nucleon decay,\\ 
\hspace*{0.3in} $\bullet$ $n-\bar n$ oscillations. 

Bounds from these have been reviewed several
times in the literature\cite{Dreiner,French-group,Russians} and will not
be discussed here. A summary of some of these
is depicted in Figure 2. These have been taken from Ref.\cite{Dreiner},
and some of them are a little outdated: some updates may be found
in Refs.~\cite{French-group} and \cite{Russians}.

\begin{figure}[h]
\begin{center}
\vspace*{2.1in}
      \relax\noindent\hskip -3.20in\relax{\includegraphics{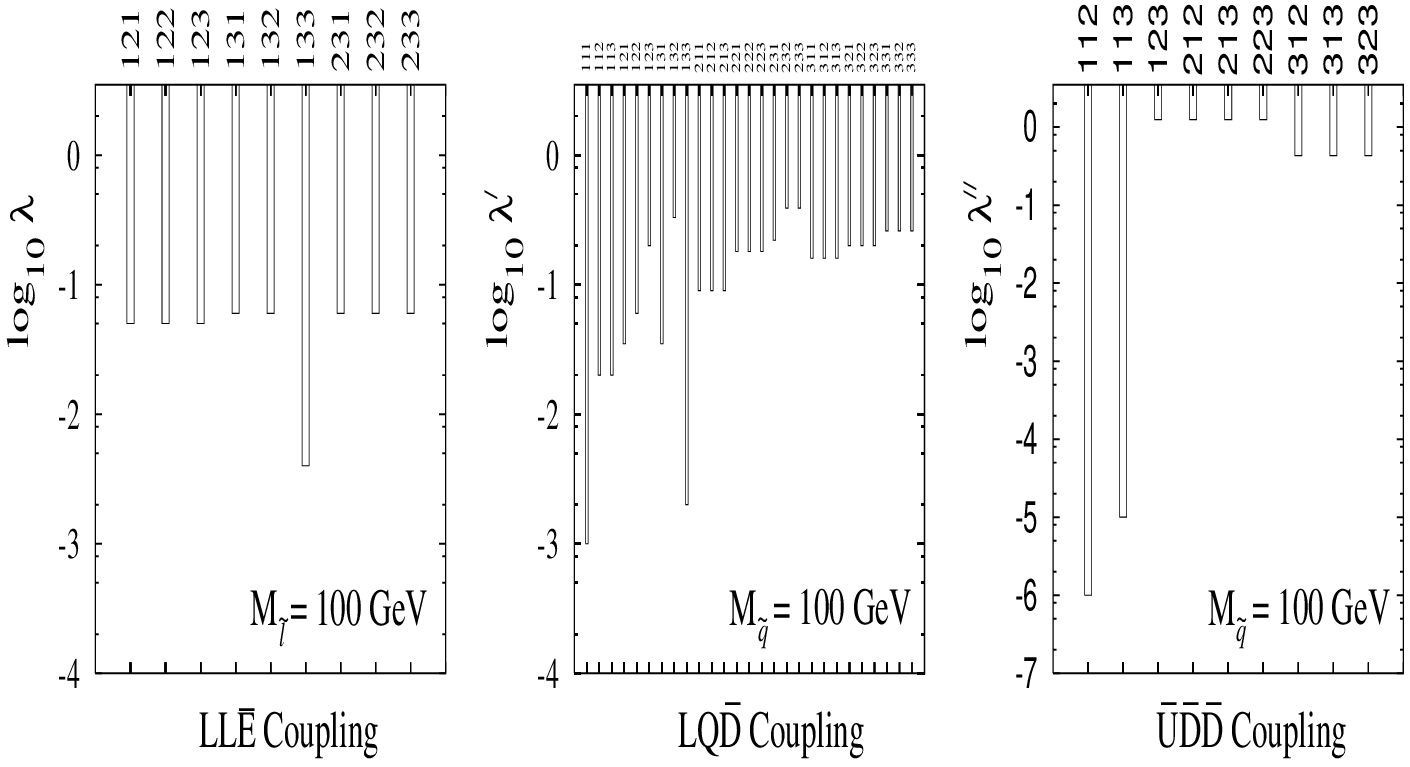}}
\end{center}
\end{figure}
\vskip 43pt
\noindent {\bf Figure 2.} {\footnotesize\it 
Bounds on $R$-parity-violating couplings\cite{Dreiner} assuming
the mass of the exchanged sfermion is 100 GeV. The vertical bars correspond
to ruled-out regions.}
\vskip 5pt

A glance at the figure will immediately reveal that bounds on couplings
involving the first
generation are generally the best, weakening as one goes through the second
generation to the third generation. Accordingly, one can predict the
greatest deviations from the SM in processes involving the third generation.
Of course, these effects are generally harder to measure, which is why the 
bounds are weak in the first place. 

It is also worth pointing out that these bounds are, in a sense,
unrealistic in the sense that it is assumed, while deriving them,
that there is no other supersymmetric contribution to the process(es)
under consideration. This would be justified if the sparticle exchanged
in an $R$-parity-violating interaction was the only light one, others
being heavy enough to decouple. Such an assumption is rather artificial.
A complete study would, however, involve all the parameters of the MSSM
and render the analysis quite messy. Preliminary 
investigations\cite{Kundu-Raychaudhuri} show, however, 
that many of the existing bounds are actually quite robust. 

The recent confirmation of an atmospheric neutrino deficit from the
experiments conducted by the Super-Kamiokande Collaboration
has led to an explosion in the 
literature of speculations about finite neutrino masses and large-angle 
mixing.  $R$-parity violation is one mechanism which can generate these,
and, in fact, the rather stringent bound on $\lambda_{133}$ comes
from a consideration of the $\nu_e$ mass\cite{Hall-Suzuki}.
For further discussions of this extremely 
interesting aspect of $R$-parity violation
the reader is referred to the literature\cite{neutrino}. \\

\centerline{\large\sc 3. Collider Signals for R-Parity Violation} 
\bigskip

In models where $R$-parity is conserved, the LSP is a
stable particle which escapes detection at colliders and leads,
as we have explained above, to large 
missing energy and momentum signatures. If $R$-parity is violated, this 
is no longer true. The LSP can
decay to multi-fermion states through the $R$-parity-violating couplings.
Since this is the sole decay mode of the LSP, the actual magnitude of the 
coupling is not very important, so long as it allows the LSP to decay
within the detector. 

Of greater importance, once we allow $R$-parity to be broken, is the
{\it identification} of the LSP. Since it decays and is therefore useless as a dark 
matter candidate, all the phenomenological arguments in favour of its being 
a sneutrino or neutralino are no longer valid. In fact,
in the most general $R$-parity-violating 
scenario, {\it any} of the sparticles can be the LSP.
One would, in general, have to construct separate search strategies for each
case. It is usual, therefore, to invoke a specific pattern of
SUSY-breaking in order to predict the mass spectrum --- this
enables us to predict which particle is the LSP. A popular choice of model is the 
{\it constrained} MSSM, or $c$MSSM, which assumes a common scalar 
mass~($m_0$), a common
gaugino mass~($M_{1/2}$) and common trilinear couplings~($A$) 
at some high scale, usually
identified with the GUT scale. In this scenario, one
possibility --- valid over a large part of the MSSM parameter space
--- is for the lightest neutralino to be the LSP, just
 as it is in the case when $R$-parity is conserved. 
Another possibility is for the LSP to be one of the sleptons. 
Most phenomenological studies and experimental searches concentrate
--- for obvious reasons --- 
on the former scenario. The latter scenario(s) has not been studied
in similar detail, though they have been considered\footnote{An example 
of this is the explanation\cite{Carena-Wagner-Giudice-Lola} 
of the excess in 
four-jet events claimed by the ALEPH Collaboration in 1995. Some explanations
of the excess in high-$Q^2$ events seen at HERA in 1997 required a charmed
squark to be the LSP\cite{HERA_LQ}.} in specific contexts. 

A slepton LSP can decay directly into two leptons or two quarks if
$L$ is violated; if $B$ is violated, it will have
four-body decays into a lepton and three quarks. On the other hand,
a neutralino LSP will always have three-body decays of the 
form\cite{Baltz-Gondolo} 
\begin{equation}
\widetilde{\chi}^0_1  \longrightarrow^{\!\!\!\!\!\!\!\!\lambda}  
~~\ell^+_1 \ell^-_2 + E_T\!\!\!\!\!\!\!/  \ \ \ , \qquad
\widetilde{\chi}^0_1  \longrightarrow^{\!\!\!\!\!\!\!\!\lambda'}  
~~\ell^\pm + 2~{\rm jets} \ , \qquad 
\widetilde{\chi}^0_1  \longrightarrow^{\!\!\!\!\!\!\!\!\lambda''}  
3~{\rm jets} \ ,
\end{equation}
and these will lead to rather distinctive final states at colliders.

Signals for $R$-parity violation at colliders can, therefore,
be classified into two kinds:
\begin{itemize}
\item where the magnitude of the 
coupling is important, such as, {\it e.g.}, virtual
sparticle exchanges and single sparticle production; 
and 
\item where the magnitude is irrelevant, {\it e.g.} in LSP decay. 
\end{itemize}
Since, at the present time, there exist no experimental 
signs of physics beyond the SM, what one can obtain from
an analysis of the first kind of signal are bounds on the $R$-parity-violating
couplings modulo assumptions about the other parameters of the MSSM.
On the other hand, the second kind of coupling leads to bounds on the
MSSM parameter space {\it in the presence of} $R$-parity violation,
without giving much information on the magnitude of the coupling
involved. 
Most current studies at colliders have been of the latter type;
studies of the former kind, though not entirely neglected, were really
triggered-off by the observation of a possible leptoquark signal at HERA 
in 1997 (see 3.3 below). However, similar studies are possible at almost
all types of colliders, and some of these will be discussed below. \\

{\sl 3.1 ~~~\underline{R-Parity Violation at LEP}} \\

At the CERN Large Electron Positron (LEP) collider three kinds of signals
are possible. Typical diagrams for these are shown in Figure 3, though
the list is far from exhaustive. The first two involve direct values
of the couplings, while the last involves decay of the LSP, which is
also an end-product of decays of the chargino.

\vskip -25pt
\hskip  15pt
\thicklines
\begin{picture}(150,100)(0,0)
\setlength{\unitlength}{0.6pt}
\put(10,90){\line(1,0){150}}
\put(10, 0){\line(1,0){150}}
\multiput(80,79)(0,-10){8}{\line(0,1){5}}
\put(80,90){\circle*{6}}
\put(80, 0){\circle*{6}}
\put(10,95){\large $e^+$}
\put(10,-25){\large $e^-$}
\put(90,35){\large $\widetilde f$}
\put(150,95){\large $\ell^+$}
\put(150,-25){\large $\ell^-$}
\end{picture}

\vskip -100pt
\hskip 125pt
\thicklines
\begin{picture}(150,100)(0,0)
\setlength{\unitlength}{0.6pt}
\put(10,90){\line(1,0){150}}
\put(10, 0){\line(1,0){150}}
\multiput(80,79)(0,-10){8}{\line(0,1){5}}
\put(80,90){\circle*{6}}
\put(80, 0){\circle*{6}}
\put(10,95){\large $e^+$}
\put(10,-25){\large $e^-$}
\put(90,35){\large $\widetilde f$}
\put(150,95){\large $\ell^+$}
\put(150,-25){\large $\widetilde{\chi}^+_1$}
\end{picture}

\vskip -100pt
\hskip 250pt
\thicklines
\begin{picture}(150,100)(0,0)
\setlength{\unitlength}{0.6pt}
\put(10,90){\line(1,0){150}}
\put(10, 0){\line(1,0){150}}
\multiput(80,79)(0,-10){8}{\line(0,1){5}}
\put(80,90){\circle*{6}}
\put(80, 0){\circle*{6}}
\put(10,95){\large $e^+$}
\put(10,-25){\large $e^-$}
\put(90,35){\large $\widetilde f$}
\put(150,98){\large $\widetilde{\chi}^+_1,\widetilde{\chi}^0_1$}
\put(150,-25){\large $\widetilde{\chi}^-_1,\widetilde{\chi}^0_1$}
\end{picture}
\vskip 20pt
\noindent {\bf Figure 3.} {\footnotesize\it 
Typical Feynman diagrams leading to signals for
$R$-parity violation at LEP. }
\vskip 5pt

The first diagram on the left
shows the effect of a virtual sfermion exchange on
dilepton (diquark) production in $e^+e^-$ collisions. Sneutrino
exchanges can lead to the production of dileptons in the final
state, if the $R$-parity-violating coupling is of the $\lambda$ type,
while squark exchanges can lead to production of a dijet final state, if
the $R$-parity-violating coupling is of the $\lambda'$ type. The 
cross-section
varies as the fourth power of the relevant coupling and hence, is
rather sensitive to its value. Of course,
identification of the coupling will depend on the tagging efficiency,
which is high for leptons, reasonable for $b$-quarks, small for $c$-quarks
and non-existent for lighter quarks. $t\bar t$-production is, of course,
forbidden by kinematics. An analysis shows\cite{Choudhury} that once
the full data are available from LEP-2, one can obtain modest bounds
on the parameter space formed by the $R$-parity-violating coupling and
the mass of the exchanged sparticle.

The second diagram (in the middle) shows one possibility for single sparticle
production
at LEP-2 through $R$-parity-violating couplings. Originally discussed in
Ref.\cite{Dreiner-Lola}, a detailed study has been performed recently
in Ref.\cite{Chemtob-Moreau-1}. Other possibilities involve final states
with a neutrino and a neutralino. In this case, one has to take
into account the final states formed by chargino or neutralino decay ---
ultimately LSP decay --- in identifying signals. These amplitudes
are linear in the $R$-parity-violating coupling involved, which means
that while they are less sensitive to the magnitude of $\lambda$,
they are sensitive to possible phases in the $R$-parity-violating
couplings\cite{Chemtob-Moreau-2}. The analysis of Ref.\cite{Chemtob-Moreau-1}
shows, in general, that LEP-2 cannot really significantly improve
the bounds already existing on $R$-parity-violating couplings. It
hardly needs to be mentioned that this
is just a limitation of the energy and luminosity available at LEP-2:
we can get far better results at a 500 GeV or 1 TeV machine with 
higher luminosity. 

The third diagram on the right of Fig.~3 actually encompasses a 
large number of possibilities. One can produce any one of the 
pairs $\widetilde{\chi}^+_1 \widetilde{\chi}^-_1$ or 
$\widetilde{\chi}^0_1 \widetilde{\chi}^0_1$ or 
$\widetilde{\chi}^0_1 \widetilde{\chi}^0_2$ depending on
the point in the MSSM parameter space where the analysis is
made. The heavier chargino and neutralino states are usually
forbidden by kinematics. All these states will decay into a pair
of LSP's $\widetilde{\chi}^0_1 \widetilde{\chi}^0_1$ which can
then decay through the $R$-parity-violating couplings listed
above\cite{Godbole-Roy-Tata}. The large multitude of possible final states 
has been extensively studied by the four LEP collaborations\cite{LEP_rpV}
and has been used to constrain the MSSM parameter space.
Some of these results, from the ALEPH Collaboration, are shown in
Figure 4. 

\begin{figure}[h]
\begin{center}
\vspace*{0.8in}
      \relax\noindent\hskip -4.20in\relax{\includegraphics{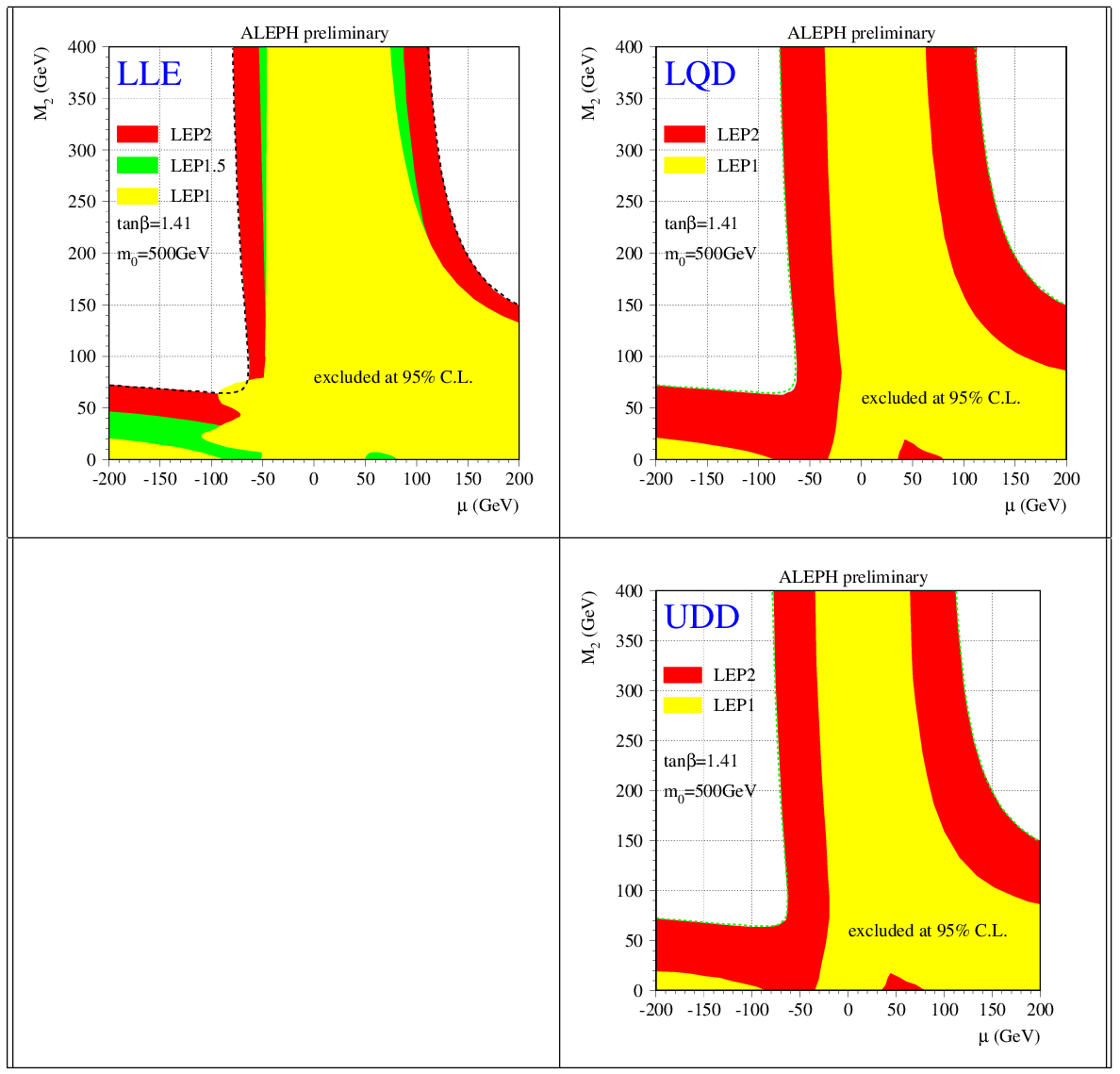}}
\end{center}
\end{figure}
\vspace*{3.5in}
\noindent {\bf Figure 4.} {\footnotesize\it
ALEPH bounds on the ($M_2,\mu$) plane for $\tan \beta \simeq \sqrt{2}$,
assuming the presence of different  $R$-parity-violating couplings. The
${\widetilde \chi}^0_1$ is assumed to be the LSP. }
\vskip 5pt 

For this study, sfermion masses were
evolved from a common $m_0 = 500$ GeV which made them too heavy to affect
the decay modes of any gaugino (other than the LSP). Moreover,
the value of $\tan\beta$ is taken to be 1.41 ($\simeq \sqrt{2}$), as indeed
it is for several other studies made by the LEP collaborations.
This is about the lowest value allowed by current LEP constraints on the plane
formed by $\tan\beta$ and the mass of the lightest Higgs boson $h^0$. Larger
values of $\tan\beta$ tend to exclude greater regions of the parameter space,
hence, the shaded regions in Fig.~4 may be taken as conservative bounds.

An important consequence of these constraints on the parameters of the
MSSM is that they can be translated into bounds on the mass of the LSP,
here assumed to
be the $\widetilde \chi^0_1$. This is because it is precisely these parameters
which go into the construction of the neutralino mass matrix.
The DELPHI Collaboration has, for example, presented
results from data collected at 183 GeV, showing that
$M_{\widetilde \chi^0_1} > 45$ GeV (unless $\tan \beta \simeq 1$), and
having an absolute lower bound of 27 GeV. For a 200 GeV run, these
numbers are expected to increase to 50-55 GeV and 35 GeV respectively.

A few more remarks may be appropriate before closing our discussion 
on LEP. As with all other areas where new physics has been predicted,
LEP has turned up negative results\footnote{
A few years back, it was thought that ALEPH had observed an excess
of events in the four-jet channel, from their 130--136 GeV data, which
appeared to come from the production of a pair of non-SM particles.
While it was shown\cite{Ghosh-Godbole-Raychaudhuri-1} that this could
not come from charginos or neutralinos decaying through $\lambda''$
couplings, a strong case was made for a pair of light 
squarks\cite{Choudhury-Chankowski-Pokorski} or 
sleptons\cite{Carena-Wagner-Giudice-Lola} decaying through
$R$-parity-violating couplings. However, a repeat run of LEP at
130-136 GeV failed to show up any further excess and hence the
four-jet `anomaly' was consigned to the dustbin of history.} 
 and hence provided a set of 
excellent (but disappointing!) bounds on the MSSM scenario where
$R$-parity is not conserved. Unless this version of supersymmetry 
is just around the corner, waiting to be discovered, the best we can
expect from the future run(s) of LEP-2 is a marginal improvement in
the bounds discussed above. \\

{\sl 3.2 ~~~\underline{R-Parity Violation at the Tevatron}} \\

At the Tevatron --- as indeed at any hadron collider ---
the second kind of signal from $R$-parity
violation can be further divided into two classes: those arising from
electroweak production of sparticles, and those arising from QCD
production of strongly-interacting sparticles, {\it i.e.} squarks and 
gluinos. For the first kind, the diagrams are analogous to those of 
Fig.~3, with the $e^+e^-$ pair being replaced by a pair of light
quarks. Typical diagrams giving rise to the production of squarks and
gluinos are shown in Fig.~5. 

\vskip 10pt 
\thicklines
\setlength{\unitlength}{0.7pt}
\begin{picture}(150,100)(0,0)
\setlength{\unitlength}{0.7pt}

\put(10,90){\line(1,0){150}}
\put(10, 0){\line(1,0){150}}

\multiput(80,79)(0,-10){8}{\line(0,1){5}}

\put(80,90){\circle*{6}}
\put(80, 0){\circle*{6}}

\put(10,95){\large $q$}
\put(10,-25){\large $\bar q$}
\put(90,35){\large $\widetilde q$}
\put(170,90){\large $\widetilde g$}
\put(170,-5){\large $\widetilde g$}

\end{picture}

\vskip -70pt 
\hskip 200pt 
\thicklines
\setlength{\unitlength}{0.7pt}
\begin{picture}(150,100)(0,0)
\setlength{\unitlength}{0.7pt}

\put(10,90){\line(1,0){80}}
\put(10, 0){\line(1,0){80}}
\put(80,90){\line(0,-1){90}}

\multiput(80,90)(10,0){8}{\line(1,0){5}}
\multiput(80, 0)(10,0){8}{\line(1,0){5}}

\put(80,90){\circle*{6}}
\put(80, 0){\circle*{6}}

\put(10,95){\large $q$}
\put(10,-25){\large $\bar q$}
\put(90,35){\large $\widetilde g$}
\put(170,90){\large $\widetilde q$}
\put(170,-5){\large $\widetilde q$}

\end{picture}
\vskip 20pt
\noindent {\bf Figure 5.} {\footnotesize\it
Typical Feynman diagrams leading to squark and gluino production 
in $q \bar q$ annihilation at a hadron collider. One can also draw
similar diagrams with initial-state gluons. }
\vskip 5pt 

Let us first consider the case of electroweak production. The
diagram analogous to the one on the extreme left of Fig.~4 can lead,
for example, to dilepton production through squark exchange. This
could lead to changes in the observed Drell-Yan cross-sections
at the Tevatron. As the observed cross-sections show remarkable
agreement with the SM ({\em d\'eja vu}) one can obtain interesting
bounds\cite{Bhat-Choudhury-Sridhar} on the parameter space from these. It
is also possible to have contributions to final states where a pair of
quarks is produced
from such diagrams (with slepton exchange). The best measured of these
cross-sections is the $t\bar t$ cross-section and it can be used
to constrain the parameter space\cite{Ghosh-Raychaudhuri-Sridhar}.

The diagrams analogous to the middle one in Fig.~3 can lead to
single sparticle production at the Tevatron. Typically in the
presence of $\lambda'$ couplings we could
expect a final state with a hard lepton and a chargino or neutralino
which would then decay in the manner described above to an LSP
and then to multi-fermion states. Analogous processes with jets could
also occur. An investigation of these possibilities is desirable,
especially in the context of Run II of the Tevatron, where reasonably
small values of the couplings could be probed.

The diagram on the extreme right of Fig.~3 will have analogues at the 
Tevatron where not only a pair of charginos or a pair of neutralinos are
produced, but a chargino can be produced in association with a neutralino. 
These, of course, are what lead to the well-known trilepton signatures.
If $R$-parity is violated, the decay of the LSP will lead to various
multi-fermion final states. For example, if there are $\lambda$-type 
couplings, one typically predicts signals with upto 7 leptons in the
final state. A detailed investigation of these is awaited. 

\begin{figure}[h]
\vspace{1.3in}
\begin{center}
      \relax\noindent\hskip -4.2in\relax{\includegraphics{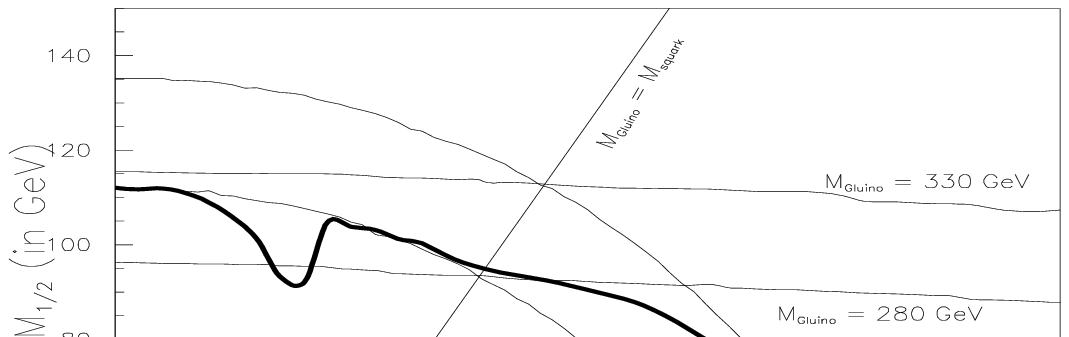}}
\end{center}
\end{figure}
\vskip 105pt
\noindent {\bf Figure 6.} {\footnotesize\it
D0 bounds on the MSSM parameter space assuming the
presence of $R$-parity-violating $\lambda'$ couplings. An $m$SUGRA
(or $c$MSSM) framework is assumed. }
\vskip 5pt

Of greater interest than the above signals, however, are those which
arise from QCD production of squarks and gluinos, which (naturally)
have far larger cross-sections. The squarks and gluinos
will decay\cite{DP-91}
through $R$-parity conserving modes (cascades), with a pair
(at least) of LSP's in the final state. When these decay, we again have a
multitude of possible final states, which have to be looked for
individually. Once again, searches for these have been performed, but
have yielded negative results so far,
leading to bounds on the MSSM parameter space. Figure 6 shows the recent
bounds made public by the D0 Collaboration\cite{D0}.
Clearly, these lead to bounds on the squark and
gluino masses over 200 GeV each and thus complement the negative
results of leptoquark searches\cite{Tev_LQ} at the Tevatron.
One can expect considerable improvement in these bounds in Run II
which essentially push the mass limits from $\sim 200$ GeV to
$\sim 300$ GeV\cite{Banerjee-etal}.

Of particular interest at the Tevatron are like-sign dilepton signals
from $\widetilde \chi_1^0$ decay in the case of $\lambda'$ couplings,
since they constitute `smoking gun' signals for Majorana fermions,
which are present in the MSSM, but not in the SM. These have been
experimentally investigated\cite{Chertok} in the context of the HERA 
anomaly (section 3.3) \
using a model described in Ref.\cite{Choudhury-Raychaudhuri-2}. 
However, a 
more general analysis has been advocated\cite{DPRoy} (with justification) 
and should be performed.

Other processes of interest at the Tevatron are the possibility of
slepton resonances\cite{Feng-etal} and the possibility of single
top quark\cite{Alakabha} and squark\cite{Berger} production. 
Though current bounds on the
couplings from these processes are not very impressive, considerable
improvement can be expected from Run-II data. 

The LEP-2 bounds should be taken into
account in studies at the Tevatron and also 
whenever future studies of $R$-parity violation are carried
out. One difficulty in the way of this is the slight difference
in philosophy between the four LEP collaborations and the CDF and D0
collaborations at Tevatron as to what constitutes the $c$MSSM, which
underlies all these studies. The CDF and D0 collaborations use the
most stringent form of the $c$MSSM, where the only free parameters
are (as explained above) $m_0$, $M_{1/2}$, $A$, $\tan\beta$
and the sign of the Higgsino mixing parameter $\mu$. The magnitude
of $\mu$ is fixed by the condition that the electrweak symmetry
be broken (by a Coleman-Weinberg-type mechanism) at the right scale.
The LEP collaborations, however, treat the magnitude of $\mu$ as
a free parameter, which essentially means that the gluino mass $M_3$
is not unified with the other gaugino masses at the GUT scale. Of
course, the gluino mass has no direct role to play at LEP, but a
comparison of LEP and Tevatron results would be possible only
if both are presented within the same set of 
assumptions\cite{Bisset-Ghosh-Raychaudhuri}.

The reverse side of the same coin is the question as to how much
these bounds can be relaxed if one does not make the assumptions
which go into the $c$MSSM. Of course, it must be admitted that in this
case there are so many free parameters that a phenomenological
analysis becomes rather bewildering. Nevertheless, it would be
interesting to know if there exist some absolute bounds from the
Tevatron and LEP-2 data on the sparticle masses. However, no
such detailed analysis exists at the present point of time.  \\

{\sl 3.3 ~~~\underline{R-Parity Violation at HERA}} \\ 

Since HERA is an $ep$ collider, it is the ideal machine to
look for first-generation leptoquarks. This was pointed out long
ago\cite{Hewett,Dreiner-Butterworth}. However, intense interest
in this subject was generated only in 1997, when the H1 and
ZEUS Collaborations simultaneously reported\cite{HERA_anomaly} 
that their data showed 
an `excess' of back-scattered (high-$Q^2$) positrons seen in 
$e^+ p$ collisions. The signals seemed to show the classic features
of resonance production of an intermediate particle which was 
quickly identified with either a leptoquark or a squark
with $\lambda'_{1j1}$ or $\lambda'_{132}$ couplings\cite{HERA_LQ}. There was 
considerable initial excitement, despite warnings\cite{Drees} 
that the H1 and ZEUS data were far from consistent with each other.
However, this euphoria was soon dissipated. In the first place, leptoquark 
searches at the Tevatron seemed to rule out a squark or leptoquark
with a mass of around 200 GeV, which was required to explain the 
HERA anomaly. Moreover, further data taken by the H1 and ZEUS 
collaborations failed to produce any further `excess' events
in the high-$Q^2$ region. Though the present data still shows
some departure from the SM, as shown in Figure 8, the
consensus appears to be that the `excess' events were due to statistical
fluctuations and do not constitute a sign for new physics beyond the SM.
In fact, both H1 and ZEUS Collaborations have now published\cite{HERA_bounds}
bounds on the MSSM parameter space analogous to those at the LEP and the
Tevatron. In the final analysis, the only positive result
of the `excess' has been to bring to common notice the fact that
$R$-parity violation is a natural SUSY scenario and not an exotic one.

\vskip 5pt
\begin{figure}[h]
\vspace{3.5in}
\begin{center}
      \relax\noindent\hskip -4.6in\relax{\includegraphics{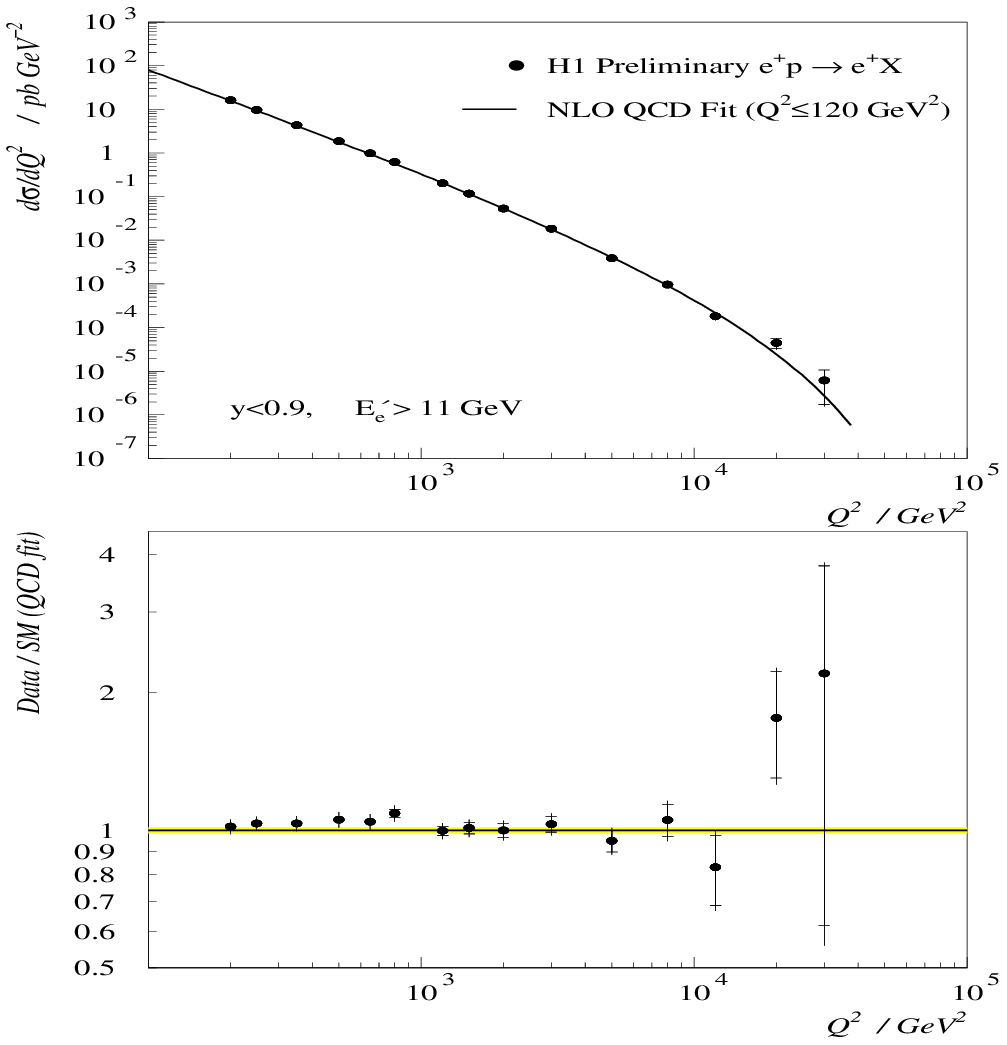}}
\end{center}
\end{figure}
\vskip 15pt
\noindent {\bf Figure 7.} {\footnotesize\it
H1 data on the $Q^2$ distribution of neutral current events. The
upper graph shows the actual distribution, while the lower one shows
the data scaled to the SM NLO prediction (the solid line in the upper 
graph). The small excess at high values of $Q^2$ is clearly within
the $2\sigma$ errors. 
}
\vskip 5pt

Several signal modes are actually possible\cite{Dreiner-Butterworth} 
at HERA, when we 
include the possibility that $R$-parity violation is weak and appears only
through decays of the LSP. Some of these involve resonance production of
a squark, as suggested in the case of the high-$Q^2$ anomaly, but there 
are other possibilities, such as associated production of a squark or 
slepton with a chargino or neutralino, followed by $R$-parity-violating
decays of both of these. The somewhat complicated signals for these have
been analyzed in Ref.\cite{Dreiner-Butterworth} and experimental searches have 
been made by the H1 Collaboration\cite{H1-old}. 
Some of the (negative) results go into the current
H1 and ZEUS bounds. An example of these is shown in Fig.~8, which 
illustrates H1 bounds on some of the $R$-parity-violating couplings
as a function of the mass of the exchanged squark. This is, in a sense,
parallelled by similar graphs produced by the LEP Collaborations.
As HERA collects more data, one can expect significant improvement in these
bounds, both in those which bound the $M_2$--$\mu$ plane, and in those
which constrain the $\lambda'$-$M_{\widetilde q}$ plane. Probably we can also
expect the $Q^2$ distribution to approximate the SM prediction better and
better as time goes by. 

\begin{figure}[h]
\vspace{3.5in}
\begin{center}
      \relax\noindent\hskip -4.6in\relax{\includegraphics{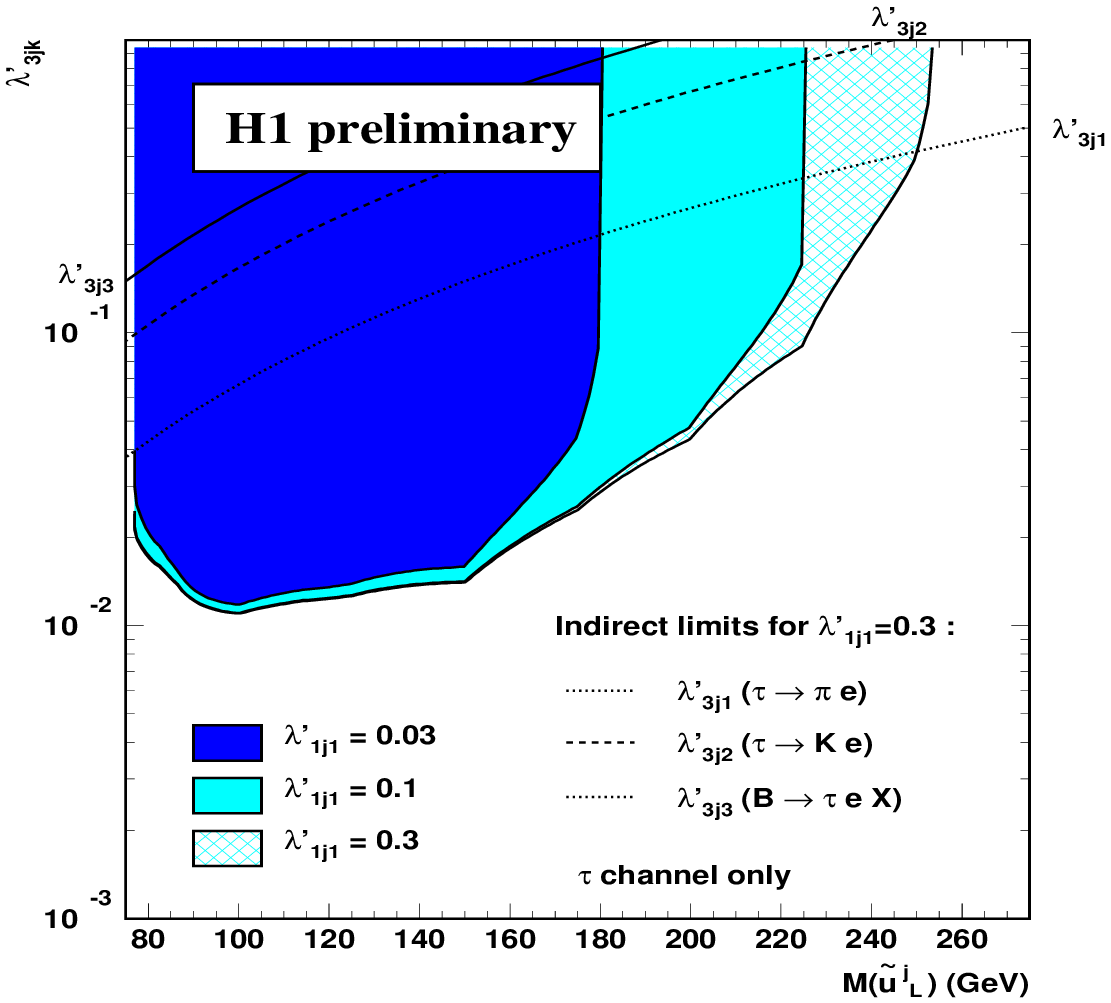}}
\end{center}
\end{figure}
\vskip 5pt
\noindent {\bf Figure 8.} {\footnotesize\it
H1 bounds on the $R$-parity violating coupling $\lambda'_{3jk}$ as a 
function of the mass of the exchanged sparticle, assuming different 
values for $\lambda'_{1j1}$. 
}
\vskip 5pt

One interesting aspect of HERA physics, which has not attracted much 
attention, is the excess of events with multiple leptons, especially muons,
in the final state. There seems to be little explanation for these in the
SM or in most beyond-SM scenarios, except as unwanted fluctuations.
While this may well turn out to be the true explanation, it is also 
possible to think\cite{Kalinowski} of a squark resonance being produced 
through a
$\lambda'$ coupling and decaying indirectly through a $\lambda$ coupling 
into multiple leptons in the final state. Such a scenario would be of 
great interest should the multi-lepton excess at HERA persist, but at
the moment it is premature to think of it as a signal for
$R$-parity-violation. \\

{\sl 3.4 ~~~\underline{R-Parity Violation at Future Colliders}} \\

Except for the Tevatron (Run-II), none of the other high energy colliders 
seem to
promise significant improvements in discovery limits for $R$-parity-violating
couplings in the near future. We must, therefore, look to the colliders of the 
future, which are expected to have much higher energy and luminosity, if
significant improvements are to be expected. The most important of these
machines is the LHC, which is scheduled to begin operation in the year 2005. 
Searches for $R$-parity violation at the LHC follow the paradigm set for
the Tevatron, the main difference being the higher energy and luminosity.
Though no {\em detailed} study of $R$-parity violating signals at the LHC 
exists
as yet, it is conceivable that some signals which are too small to be
observable at the Tevatron might become observable at the LHC. The reverse 
side of the coin is that backgrounds  --- especially QCD backgrounds ---
could also be large. However, it is possible to be upbeat about $R$-parity
violation searches at the LHC in view of the fact that 
preliminary studies carried out by the CMS and ATLAS
collaborations at the LHC\cite{LHC} promise to carry the
squark and gluino mass search limits, in the presence of
$R$-parity-violating couplings, to the vicinity of 2 TeV each.
This represents an increase of nearly an order of magnitude over the
present bounds from the Tevatron. 

\begin{figure}[h]
\vspace{3.2in}
\begin{center}
      \relax\noindent\hskip -4.6in\relax{\includegraphics{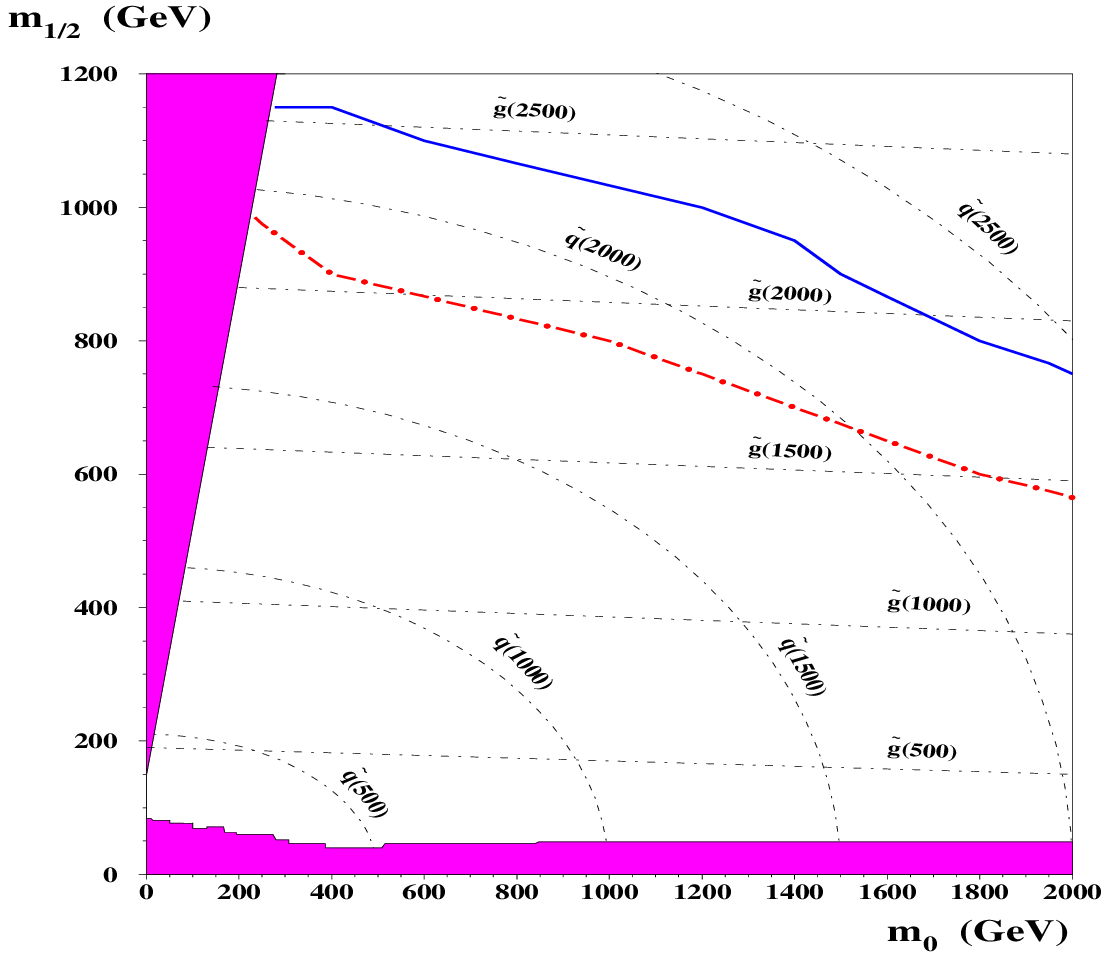}}
\end{center}
\end{figure}
\vskip 10pt
\noindent {\bf Figure 9.} {\footnotesize\it
$5\sigma$ discovery potential of CMS in mSUGRA for
$\mu < 0$, $\tan \beta = 2$
and $A_0 = 0$. In the region below the solid (blue) curve
a signal of $R$-parity-violating supersymmetry
via $\lambda_{121} = 0.05$ would be discovered
($5\sigma$ for an integrated luminosity of $10^4$ pb$^{-1}$). The 
dashed (red)
curve corresponds to the discovery potential for a signal
via $\lambda_{233} = 0.06$. In the shaded (pink) region, mSUGRA is not
valid or $\widetilde{\chi}_1^0$ is no longer the LSP.
}
\vskip 5pt

In Fig.~9, we illustrate the reach of LHC in the above parameters
using a plot from a preliminary study carried out by the CMS 
Collaboration\cite{LHC}.
In this plot contours of the squark and gluino masses are shown in the 
$m_0$--$m_{1/2}$ plane. The (pink) shaded regions show the regions
where this analysis is not valid, either because there exists no
phenomenologically viable solution to the renormalization group equations 
at the electroweak scale, or because the lightest neutralino is no 
longer the LSP. The latter case will have its own distinctive signatures
--- the fact that the relevant region is shaded merely represents
the fact that these are not covered by the analysis that generated
this plot. It is immediately apparent that for an optimistic 
$\lambda_{121} = 0.05$, where the LSP decays into multiple electron
and muon states, one can essentially probe all values of $m_{1/2}$
upto about 800 GeV, which corresponds to a gluino mass of nearly
2 TeV. The situation is worse for the $\lambda_{233}$ coupling,
despite its higher value, essentially because of the difficulties
of identifying $\tau$-leptons from the enormous backgrounds at the
LHC.

The ATLAS Collaboration has made a very similar study, basing their
signals on a $\lambda$-type coupling, where the final state contains
multi-leptons and hence is easier to detect. However, it is not clear
how efficient the LHC would be as a detector of $R$-parity violation
if the couplings are of $\lambda'$ or $\lambda''$ type. All that
can be said with certainty is that detection of the resultant signals
would be complicated by the presence of large QCD backgrounds and 
detection of like-sign dileptons might be the best bet\cite{DPRoy} 
for such studies. A detailed analysis is awaited. 
 
Next to the LHC, which is certainly going to be built, we need to consider
the Next Linear Collider (NLC), which is the usual acronym for a 500 GeV
$e^+ e^-$ collider (necessarily of the linac type). Various possibilities
for the NLC are under serious consideration. It is certainly going to be
run in the $e^+ e^-$ collision mode, but other alternatives are an
$e^- e^-$ collision mode, and, using laser back-scattering techniques,
$e \gamma$ and $\gamma \gamma$ collision modes as well. Detailed studies
of $R$-parity violation at the NLC have not really been carried out.
Preliminary studies show\cite{Dreiner-Lola, Ghosh-Godbole-Raychaudhuri-2} 
that some 
order can probably be extracted out of a multitude of signals in the 
$e^+e^-$ mode --- in fact, for $\lambda$-type couplings, we might have
rather spectacular multi-lepton signals with practically no background. 
For $\lambda'$ and $\lambda''$ couplings,
it might be possible\cite{Ghosh-Godbole-Raychaudhuri-2} 
to reconstruct the mass of the decaying LSP, assumed
to be the $\widetilde{\chi}_1^0$. 
Similar studies could also be carried out in the
$e \gamma$ mode\cite{Ghosh-Raychaudhuri}. Except for some initial 
studies\cite{Mahanta}, the $e^- e^-$ mode has not been looked into
in any detail, though this might be thought of as an ideal machine to 
study lepton number violation. Except for a preliminary study in 
Ref.~\cite{Pietschmann}, the $\gamma\gamma$ mode has not been
considered as yet for $R$-parity-violating signals. A study of this is
probably worth carrying out. 

Another possibility, which is now receiving serious consideration, is
that of a muon collider. Muons have an advantage over electrons in 
their heavier mass, which inhibits synchrotron radiation. It might,
therefore, be possible to build a $\mu^+ \mu^-$ collider with much
higher energy and luminosity than even the NLC. One important advantage
of such a machine would be in its great potential for studying 
flavour physics, of which $R$-parity violation provides one example. 
Some preliminary studies of $R$-parity violation at a muon collider
have been carried out\cite{Feng-Gunion-Han} 
and they seem to yield promising results. 

It might also be possible to have a muon-proton collider. At Fermilab
and LEP, for example, high energy proton beams are already available,
and it would only be necessary to bring these into collision with the
muon beam. Such a machine would be the second generation equivalent of 
HERA and could probe, among other things, $R$-parity-violation through
$\lambda'$ couplings to great accuracy\cite{Carena-Quigg-Raychaudhuri}. 

A muon collider may also be thought of as a copious source of neutrinos
which come from the decay of muons in the high-luminosity beams. While
it has been speculated that such neutrino beams can be used for 
long-baseline experiments to test neutrino masses and mixings, it
might also be possible, by allowing the high energy neutrino beams to
impinge on a target, to derive bounds on $R$-parity-violating couplings. 
Though similar studies have been done for ultra-energetic neutrinos
from cosmic sources\cite{Carena-Choudhury-Lola-Quigg}, the question 
of neutrinos from a muon collider has not yet been addressed. 
\\

\centerline{\large\sc 4. Summary and Outlook} 
\bigskip

$R$-parity conservation, as we have discussed, is a pleasing but not 
an absolutely
 necessary feature of the MSSM. Of course, any arguments for or against 
such a symmetry in the MSSM are essentially speculative, since,
in the first place, we have no empirical evidence that SUSY exists
--- whether in the $R$-parity-conserving or $R$-parity-violating form.
The question of $R$-parity should therefore be
approached with an open mind. What is known
for certain is that there are useful bounds from low-energy
studies and from collider experiments which are still running or
will run again in the near future. Of these, bounds from LEP and
the Tevatron are essentially indirect bounds, in that they arise
from a search for LSP decay modes. More direct bounds come from HERA,
which, after generating some initial excitement, is now expected 
to provide steadily improving bounds, as more and more data is
accumulated. Major improvements may be expected from Run-II of the
Tevatron, though a hadron machine has intrinsic problems when
dealing with multi-jet final state signals. It is, therefore, 
necessary to 
continue investigations of possible signals for $R$-parity violation
not only at the present generation of colliders, but also at
the machines of the future, of which the Large Hadron Collider (LHC)
at CERN is a certainty. Less certain, but quite probable, are linear 
$e^+ e^-$ colliders while muon colliders are now being seriously
considered. A major part of the physics agenda at
such machines would be SUSY searches and hence, studies 
of $R$-parity violation are essential to have a comprehensive picture.

It must be noted that 
present work in $R$-parity violation has merely touched the
tip of the iceberg, since it has usually been carried out
within a very specific set of assumptions --- such as, {\it e.g.} 
a neutralino LSP. Apart from studies connected with experimental searches,
most studies at future colliders have been rather sketchy in addressing
issues such as, for example, QCD corrections and backgrounds, both from 
the SM as well as from the $R$-parity-conserving sector of the MSSM. 
In fact, till now, studies of $R$-parity violation have usually
assumed that the rest of the MSSM does not matter, an assumption which
can, with justification, be challenged. 
The future will probably see a number of the issues
discussed here 
being addressed and, with a bit of luck, the world may turn out
to be supersymmetric and $R$-parity-violating as well. \\

\centerline{\large\bf Acknowledgements} 
\bigskip

The author would like to thank Prof.~V.~Narasimham and the organisers 
of the 13th Topical Conference on Hadron Collider Physics at Mumbai
(January 1999) for the opportunity to deliver this talk. 
He would also like to thank his various collaborators,
especially Debajyoti Choudhury, Rohini M. Godbole and Dilip Kumar Ghosh, 
who have been mostly instrumental in introducing him to the subject.

\end{document}